\documentclass[aps,showpacs,superscriptaddress]{revtex4}
\usepackage{latexsym,amsmath,amssymb,epsfig}
\newcommand{\bea}{\begin{eqnarray}}
\newcommand{\eea}{\end{eqnarray}}
\newcommand{\be}{\begin{equation}}
\newcommand{\ee}{\end{equation}}
\newcommand{\beast}{\begin{eqnarray*}}
\newcommand{\eeast}{\end{eqnarray*}}
\newcommand{\pkt}{\; .}
\newcommand{\kma}{\; ,}
\newcommand{\nn}{\nonumber}

\newcommand{\labelcaption}[2]{\caption[#1]{\label{#1}#2}}
\def\e{{\rm e}}

\newcounter{subequation}[equation]
\makeatletter
\expandafter\let\expandafter\reset@font\csname reset@font\endcsname
\newenvironment{subeqnarray}
  {\arraycolsep1pt
    \def\@eqnnum\stepcounter##1{\stepcounter{subequation}{\reset@font\rm
      (\theequation\alph{subequation})}}\eqnarray}%
  {\endeqnarray\stepcounter{equation}}
\makeatother

\begin{document}
%

\rightline{AEI-2010-028}

\title{On instability of Rubakov-Shaposhnikov model}
\author{George Lavrelashvili \footnote{lavrela@itp.unibe.ch }}
\affiliation{
Max-Planck-Insitut f\"ur Gravitationsphysik, Albert-Einstein-Institut \\
Am M\"uhlenberg 1, DE-14476 Potsdam, Germany}
\affiliation{Department of Theoretical Physics,
A.Razmadze Mathematical Institute \\
GE-0193 Tbilisi, Georgia}
\date{\today}

\begin{abstract}
\noindent
Instability of 6 dimensional Rubakov-Shaposhnikov model
is reinvestigated. It is shown that the model is unstable in scalar perturbations sector
with very particular instability pattern: there are no unstable modes for the first two lowest
angular harmonics, $m=0$ and $m=1$, whereas there is a single negative mode for each higher $m$.
\end{abstract}
\maketitle

\section{Introduction}

In the beginning of 80th two very important papers by Rubakov and Shaposhnikov were published
in the same issue of Physics Letter B \cite{rs83a}, \cite{rs83b}.
The first one \cite{rs83a} was discussing possibility that (in modern language)
we live on a brane in higher dimensional space and in the second one \cite{rs83b}
the warped compactification was introduced in order to attack cosmological constant
problem. These ideas created basis for ``extra dimensional revolution"
 which happen 15 years later \cite{dvali98a},
\cite{dvali98b}, \cite{dvali98c}, \cite{gogber98}, \cite{ransun99a}, \cite{ransun99b}.

The stability of the Rubakov-Shaposhnikov model with warped compactification
\cite{rs83b} was questioned \cite{lt85} soon after the model was suggested.
It was found that the model is stable under tensor and vector perturbations, but
has unstable modes in scalar perturbations sector. Recently we became aware
\footnote{We are grateful to Zurab Ratiani for pointing out this error.}
that there is an algebraic error in the prove namely in the Eq.(23) of \cite{lt85}.
The aim of present note is to to correct this error and reinvestigate the stability of
the Rubakov-Shaposhnikov model.
Since in addition there are numerous misprints in the most of
equations in the journal version of \cite{lt85},
first we repeat here derivation of Schr\"oedinger equations and then
give direct numerical proof of existence of an unstable modes
in scalar perturbations sector.

\section{Spontaneous compactification with zero cosmological constant} \label{model}
\par
The solution, leading to zero four dimensional cosmological constant,
proposed in \cite{rs83b} looks as follows.
We consider gravity in $d+N-$dimensional
space-time with the metric $\hat{g}_{AB}$ (signature $+ - ... -$). The Einstein equations is written
with the cosmological constant:
\be \label{eq1}
\hat{R}_{AB}-\frac{1}{2}\hat{g}_{AB} \hat{R} = \Lambda \hat{g}_{AB}\pkt
\ee
It is assumed that $\Lambda>0$. With the warped ansatz for the metric
\be \label{eq2}
\hat{g}_{AB}=
\left(
\begin{array}{cc}
\sigma(x^a) g_{\mu\nu} (x^\lambda) & 0 \\
0  & \tilde{g}_{ab} (x^a)
\end{array}
\right) \kma
\ee
the Eq.~(\ref{eq1}) reduces to the system of equations
\bea
&&R_{\mu\nu}-\frac{1}{2}g_{\mu\nu}R=\Lambda_{\rm phys} g_{\mu\nu} \kma \label{eq3}\\
&&\tilde{R}_{ab}=-\frac{2}{N+d-2} \Lambda \tilde{g}_{ab}
+d (\frac{\tilde{\nabla}_a \tilde{\nabla}_b \sigma}{2 \sigma}
-\frac{\tilde{\nabla}_a \sigma \tilde{\nabla}_b \sigma}{4 \sigma^2}) \kma \label{eq4}\\
&&\frac{1}{2}\tilde{\nabla}_a\tilde{\nabla}^a \sigma
+(d-2) \frac{\tilde{\nabla}_a \sigma \tilde{\nabla}^a \sigma}{4 \sigma}
-\frac{2 \Lambda}{ N+d-2} \sigma = - \frac{2 \Lambda_{\rm phys}}{d-2} \pkt \label{eq5}
\eea
the hats and tildas respectively denote $d+N$-dimensional and $N$-dimensional
quantities, $\mu, \nu, ... = 0,1,...,d-1$, and $a,b,... = d,...,d+N-1$.

In these equations $\Lambda_{\rm phys}$ is an arbitrary parameter arising from the separation
of variables, $R_{\mu\nu}$ is constructed from $g_{\mu\nu}$ and and $\tilde{R}_{ab}$ and $\tilde{\nabla}_a$
are constructed from $\tilde{g}_{ab}$ according to the usual rules.
The Latin indices $a,b,...$ are raised and lowered  with the metric $\tilde{g}_{ab}$.
Note that, in contrast to the standard approach to spontaneous compactification, the space defined
by the metric $\hat{g}_{AB} $ is not the direct product of the $d$- and $N$-dimensional spaces.
This difference is related to the presence of a warped factor $\sigma (x^a )$ in front of
$g_{\mu\nu} (x)$ and is critical future in the entire discussion.

The Eq.~(\ref{eq3}) is the Einstein equation for the $d$-dimensional metric $g_{\mu\nu}$ with the
cosmological constant $\Lambda_{\rm phys}$. For $\Lambda_{\rm phys}=0$ it has a solution corresponding
to a flat space. Assuming $d=4$ and $N=2$, the
equations (\ref{eq4}) and (\ref{eq5}) can be solved with the result:
\bea
\tilde{g}_{ab}&=& \left( \begin{array}{cc}
-1 & 0 \\
0  & f(\rho) \end{array} \right) \kma  \label{eq6} \\
f(\rho)&=&-\frac{8}{5\Lambda} \left[{\rm tg}(\sqrt{\frac{5\Lambda}{8}} \rho)\right]^2
\left[{\rm cos}(\sqrt{\frac{5\Lambda}{8}} \rho)\right]^{4/5} \label{eq7} \\
\sigma(\rho) &=& \left[{\rm cos}(\sqrt{\frac{5\Lambda}{8}} \rho)\right]^{4/5} \label{eq8} \kma
\eea
where $\Lambda_{\rm phys}=0, x^4\equiv \rho, \rho \in [0,\rho_{\rm max}],
x^5 \equiv \theta, \theta\in [0, 2 \pi]$.

In spite the fact that this solution is noncompact in the usual sense
(the circumference $x^\mu={\rm const}, \rho={\rm const}$ can be arbitrarily large,
$l=2\pi \sqrt{-f}\to\infty$ as $\rho\to\rho_{\rm max}$), it can be shown that the presence
of the two extra dimensions is unobservable at low energies \cite{rs83b}.

\section{STABILITY ANALYSIS} \label{stab}
\par

\subsection{The equations of motion} \label{eqs}
\par
Let us find the equations of motion for a small fluctuations about the solution $\hat{g}_{AB}^0$.
For this we substitute $\hat{g}_{AB}\to\hat{g}_{AB}^0+\epsilon_{AB}$ into the Einstein equations
Eq.~(\ref{eq1}). Neglecting all terms with powers higher than first in $\epsilon_{AB}$, we find
\be \label{eq9}
\hat{\nabla}_C \hat{\nabla}_A {\epsilon_B}^C
+\hat{\nabla}_C \hat{\nabla}_B {\epsilon_A}^C
-\hat{\nabla}_C \hat{\nabla}^C \epsilon_{AB}
-\hat{\nabla}_A \hat{\nabla}_B {\epsilon_C}^C
=\frac{4 \Lambda}{2-N-d} \epsilon_{AB} \kma
\ee
where the covariant derivatives $\hat{\nabla}_A$ are calculated using the background
metric $\hat{g}_{AB}^0$. In what follows we set $d=4$ and $N=2$.
Next we use the fact that $\hat{g}_{AB}^0$
is independent of $x_\mu$ and go to the momentum representation in $x_\mu$. Denoting
\bea \label{eq10}
\epsilon_{\mu\nu}=\sigma h_{\mu\nu},~\quad {\epsilon_{\mu}}^\nu= {h_{\mu}}^\nu,~\quad
\epsilon^{\mu\nu}=\sigma^{-1} h^{\mu\nu}\kma \nn \\
\epsilon_{a\mu}=A_{a\mu},~\quad {\epsilon_a}^\mu= \sigma^{-1} {A_a}^{\mu},~\quad
\epsilon_{ab}=\varphi_{ab} \kma
\eea
and substituting the decomposition of $h_{\mu\nu}$ and $A_{a\mu}$  into components with
spin $0,1$ and $2$
\bea \label{eq11}
h_{\mu\nu} (k,x^a)=l_{\mu\nu}+k_\mu f_\nu +k_\nu f_\mu +\frac{k_\mu k_\nu}{k^2} P
+(\eta_{\mu\nu}- \frac{k_\mu k_\nu}{k^2}) S \kma \nn \\
A_{a\mu} (k,x^a) = r_{a\mu} +k_\mu \Phi_a \kma
\eea
where
\be
k^\mu l_{\mu\nu}=0,~\quad k^\mu r_{a\mu}=0,~\quad k^\mu f_\mu=0,~\quad {l_\mu}^\mu=0 \kma
\ee
from the Eq.(\ref{eq9}) we obtain seven separate equations for tensor, vector and scalar
(under rotations of $x_\mu$) perturbations:
\be \label{tensor_eq}
k^2 \sigma^{-1} l_{\mu\nu}-(\tilde{\nabla}_a \tilde{\nabla}^a
+2 \pi_a \tilde{\nabla}^a ) l_{\mu\nu}=0 \; ;
\ee
\begin{subeqnarray}\label{vector_eq}
i (\tilde{\nabla}^a+\pi^a) r_{a\nu}
-\sigma (\tilde{\nabla}_a \tilde{\nabla}^a +2\pi_a  \tilde{\nabla}^a) f_\nu &=& 0 \kma \\
k^2 \sigma^{-1}  r_{a\mu}+i k^2 \tilde{\nabla}_a f_\mu
-(\tilde{\nabla}_b \tilde{\nabla}^b+\pi_b  \tilde{\nabla}^b -\Lambda )r_{a\mu}
+(\tilde{\nabla}_b \tilde{\nabla}_a -\pi_a \tilde{\nabla}_b + 2 \pi_b \tilde{\nabla}_a
-\tilde{\nabla}_b \pi_a - 2 \pi_a \pi_b) {r_\mu}^b
 &=& 0 \; ;
\end{subeqnarray}
\begin{subeqnarray}\label{scalar_eq}
k^2 \sigma^{-1}S +i k^2 \sigma^{-1} \pi_a \Phi^a - \frac{1}{2} \pi_a \tilde{\nabla}^a
(P+3 S+{\varphi_b}^b) +(2\pi^a \pi^b+\tilde{\nabla}^a \pi^b+\pi^b \tilde{\nabla}^a ) \varphi_{ab}
-(\tilde{\nabla}_a \tilde{\nabla}^a+2\pi_a \tilde{\nabla}^a ) S =0  \kma \\
2 i k^2 \sigma^{-1}(\tilde{\nabla}^a +\pi^a) \Phi_a+ k^2 \sigma^{-1} (2 S+ {\varphi_b}^b)
-(\tilde{\nabla}_a \tilde{\nabla}^a+2\pi_a \tilde{\nabla}^a)(P-S)= 0 \kma \\
i ( \tilde{\nabla}_b {\varphi_a}^b - \tilde{\nabla}_a (3 S +{\varphi_b}^b)+\frac{1}{2}\pi_a {\varphi_b}^b
+\pi_b {\varphi_a}^b)
-(\tilde{\nabla}_b  \tilde{\nabla}^b+2 \pi_b \tilde{\nabla}^b-\Lambda )\Phi_a \nn \\
+ (\tilde{\nabla}_b \tilde{\nabla}_a -\pi_a \tilde{\nabla}_b +2 \pi_b \tilde{\nabla}_a
-\tilde{\nabla}_b\pi_a - 2 \pi_a\pi_b )\Phi^b = 0 \kma \\
k^2 \sigma^{-1}\varphi_{ab}+i k^2 \sigma^{-1} (\tilde{\nabla}_a \Phi_b+ \tilde{\nabla}_b \Phi_a )
-\frac{1}{2}(\pi_b \tilde{\nabla}_a +\pi_a \tilde{\nabla}_b)(P+3S)
-\tilde{\nabla}_a \tilde{\nabla}_b (P+3 S) \nn \\
- \tilde{\nabla}_a \tilde{\nabla}_b {\varphi_c}^c
+(\tilde{\nabla}_c +2 \pi_c)(\tilde{\nabla}_a {\varphi_b}^c + \tilde{\nabla}_b {\varphi_a}^c )
+ (\Lambda- \tilde{\nabla}_c \tilde{\nabla}^c-2 \pi_c \tilde{\nabla}^c) \varphi_{ab} = 0 \kma
\end{subeqnarray}
where $\pi_a = (\partial/\partial x^a) {\rm ln} \sigma$.

The system of equations (\ref{tensor_eq}-\ref{scalar_eq}) is a system of eigenvalue equations
with the role of the unknown eigenvalue played by $k^2$.
If a system (for example, (\ref{scalar_eq})) is consistent,
there will be at least one equation of the form
\be \label{schr_eq}
L \Psi (k, x^a)= k^2 \Psi (k, x^a) \kma
\ee
for some $\Psi (k, x^a)$, where $L$ is a differential operator acting on $x^a$.
The question of stability of the solution $\hat{g}_{AB}^0$ now reduces to the question of the
possible values of $k^2$.
The solution will be linearly stable if $k^2$ has no negative values and solution is unstable
if $k^2$ can take negative values.

Note that in this problem there is gauge invariance related to the invariance of the equation (\ref{eq1})
under general coordinate transformations $x^A \to {x'}^A=x^A - \Delta x^A$. In terms of variables
Eq.~(\ref{eq10}) and Eq.~(\ref{eq11}), gauge transformations with the parameters
$\Delta x^A = [\sigma^{-1} (\omega^{\mu}+k^{\mu} \eta), \chi^a]$ looks like
\bea\label{gauge_trans}
\delta l_{\mu\nu}=0 ,~~~ \delta r_{a\mu}= (\frac{\partial}{\partial x^a}-\pi_a)\omega_\mu ,~~~
\delta f_\mu = i \sigma^{-1} \omega_{\mu} ,~~~
\delta \Phi_a= (\frac{\partial}{\partial x^a}-\pi_a)\eta+i\chi_a , \nn \\
\delta S=\pi_a \chi^a ,~~~\delta P =2i k^2 \sigma^{-1} \eta +\pi_a \chi^a ,~~~
\delta\varphi_{ab}=
(\frac{\partial \tilde{g}_{ab}}{\partial x^c}
+\tilde{g}_{ac} \frac{\partial }{\partial x^b}
+\tilde{g}_{bc}\frac{\partial }{\partial x^a}) \chi^c .
\eea

\subsection{The mass spectrum} \label{spectrum}
\par

It follows directly from the Eq.~(\ref{tensor_eq}) for tensor perturbations that
$k^2$ is non-negative for this sector.
In order to see this, we set $\mu, \nu=0$ in this equation, multiply both sides by
$\sigma^2 \sqrt{-f} l^{00}(-k)=[-det(\hat{g}^0_{AB})]^{1/2} l^{00} (-k)$, and integrate over
$dx^4 dx^5 \equiv d\tilde{x}$. Integrating the right hand side by parts
(the correctness of this procedure can be rigorously justified), we obtain the equation
\be
k^2 \int \sigma \sqrt{-f}d\tilde{x} |l_{00} (k)|^{2} =- \int \sigma^2 \sqrt{-f} d\tilde{x}
[\tilde{\nabla}_a l_{00} (k)]^{\ast}\; [\tilde{\nabla}^a l^{00}(k)] \kma
\ee
from which it follows that $k^2\geq 0$.

For vector perturbations first we have to fix the gauge.
We choose the gauge conditions as
\be \label{gc_vector}
f_{\mu}=0 \pkt
\ee
Furthermore, we set $\mu=0$ in the second of equations (\ref{vector_eq}), multiply
by $\sigma \sqrt{-f} r^{a0} (-k)$, sum over $a$, and integrate over $d\tilde{x}$.
Then with the help of gauge condition Eq.~(\ref{gc_vector}) and use of the first of equation
(\ref{vector_eq}) after some transformations we find
\be
k^2 \int \sqrt{-f} d\tilde{x} r^{a0} r_{a0} =\int \sigma \sqrt{-f} d\tilde{x}
(-\tilde{\nabla}_a r_{b0} \tilde{\nabla}^a r^{b0}-\frac{1}{2}
\pi_a \pi_b r^{a0} r^b_0-\frac{3}{4}\pi_c \pi^c r^{a0} r_{a0} ) \pkt
\ee
The integral multiplying $k^2$ and the right hand side are both non-positive.
Therefore, for the vector perturbations also $k^2 \geq 0$.

Let us now turn to equations Eq.~(\ref{scalar_eq}) for scalar perturbations.
We choose the gauge condition in the form
\be \label{gc_scalar}
\Phi_a=0,~\quad P-S=0 \pkt
\ee
We shall assume that $k^2\neq 0$. Then equation (\ref{scalar_eq}a) is a consequence
of the three other equations and can be omitted. From Eq.~(\ref{scalar_eq}b) we obtain
\be
S=-\frac{1}{2} {\varphi_b}^b \kma
\ee
which can be taken as a definition of $S$ in terms of ${\varphi_a}^b$.
Using this equation and the gauge condition Eq.~(\ref{gc_scalar}), the two remaining
equations can be written in terms of ${\varphi_a}^b$:
\begin{subeqnarray}\label{scalar_eq_2}
&&(\tilde{\nabla}_a + \pi_a) {\varphi_b}^b = -2 (\tilde{\nabla}_b + \pi_b) {\varphi_a}^b   \kma \\
&&k^2 \sigma^{-1} \varphi_{ab}=
-(\tilde{\nabla}_a \tilde{\nabla}_b+\pi_a \tilde{\nabla}_b+\pi_b \tilde{\nabla}_a ){\varphi_c}^c
-(\tilde{\nabla}_c+2 \pi_c)(\tilde{\nabla}_a {\varphi_b}^c +\tilde{\nabla}_b {\varphi_a}^c )
+(\tilde{\nabla}_c \tilde{\nabla}^c + 2 \pi_c \tilde{\nabla}^c - \Lambda)\varphi_{ab}  \pkt
\end{subeqnarray}
Equation (\ref{scalar_eq_2}a) gives two relations between three variables $\varphi_{ab}$,
so a single independent variable remains. Three equations (\ref{scalar_eq_2}b) are equivalent
to each other and determine the spectrum of $k^2$. The problem is to solve the constraint
(\ref{scalar_eq_2}a), that is, to express all the in terms of a single independent variable $\xi$
and its derivatives. Then from (\ref{scalar_eq_2}b) we find following equation for $\xi$:
\be
k^2 \xi = M \xi \kma
\ee
where $M$ is a differential operator acting on $x^a$.

Let us expand $\varphi_{ab} (k, \rho, \theta)$ in a Fourier series in $\theta$
\be
\varphi_{ab}(\rho, \theta) = \sum_{m=-\infty}^{\infty} \e^{im\theta} \varphi_{ab m} (\rho)
\kma
\ee
and consider cases $m=0$ and $m \neq 0$ separately.

For $m=0$ the Eq.~(\ref{scalar_eq_2}a) gives:
\begin{subeqnarray}\label{eq20}
(\partial_{\rho}+\Gamma +\pi_4) {\varphi_5}^4=0 \kma \\
(\partial_{\rho}+\pi_4)(3{\varphi_4}^4+{\varphi_5}^5)=
-2 \Gamma ({\varphi_4}^4 -{\varphi_5}^5) \kma
\end{subeqnarray}
where $\Gamma=\tilde{\Gamma}^5_{45}=\frac{1}{2} \partial_\rho {\rm ln} f$.
The first of these equations has no regular solutions except for zero, so it gives
${\varphi_5}^4=0$. Using this condition, the Eq.~(\ref{eq20}b), the background equation (\ref{eq5})
and denoting $ 3 {\varphi_4}^4+ {\varphi_5}^5 \equiv \xi_0 $, we find
\be
k^2 \sigma^{-1} \xi_0 = -\partial_{\rho}^2 \xi_0
+(2 \frac{\partial_{\rho}\Gamma}{\Gamma}-\Gamma-\pi_4 ) \partial_{\rho} \xi_0
+(2 \pi_4 \frac{\partial_{\rho}\Gamma}{\Gamma}-2 \pi_4 \Gamma-\pi_4^2 ) \xi_0 \pkt
\ee
Multiplying this equation by $\xi_0$ and integrating over $\rho$ from zero to $\rho_{\rm max}$
with the weight $W(\rho)$ given by the expression
\be
W(\rho)={\rm exp}\left( -\int d\rho (2 \frac{\partial_{\rho}\Gamma}{\Gamma}-\Gamma-\pi_4 )
\right) \geq 0 \kma
\ee
after integration by parts we obtain
\be
k^2 \int \sigma^{-1}W \xi_0^2 d\rho = \int W d\rho \left[(\partial_{\rho}\xi_0)^2
+ (2 \pi_4 \frac{\partial_{\rho}\Gamma}{\Gamma}-2 \pi_4 \Gamma-\pi_4^2 )\xi_0^2 \right] \pkt
\ee
The positivity of the last term in the integrand follows from the explicit form of
background solution, (\ref{eq7}) and (\ref{eq8}). Therefore, $k^2\geq 0$ for $m=0$.

In the $m\neq 0$ case we denote
$ {{\tilde\varphi}_{5~m}}^{~4} (\rho)= \frac{1}{im} {\varphi}_{5~m}^{~4} (\rho)$.
Dropping the index $m$ on the fields $\varphi_{ab}$,
we rewrite Eq.~(\ref{scalar_eq_2}a) as
\begin{subeqnarray}
(\partial_\rho+\pi_4) (3{\varphi_4}^4+{\varphi_5}^5)=2\mu {{\tilde\varphi}_{5}}^{~4}
-2 \Gamma ({\varphi_4}^4-{\varphi_5}^5) \kma  \\
{\varphi_4}^4+ 3 {\varphi_5}^5= -2(\partial_\rho+\Gamma+\pi_4) {{\tilde\varphi}_{5}}^{~4} \kma
\end{subeqnarray}
where $\mu=-m^2/f$. The solution of this system has the form:
\bea
\xi\equiv 3{\varphi_4}^4+{\varphi_5}^5 +2 \Gamma {{\tilde\varphi}_{5}}^{~4} \kma~\quad
G \equiv \mu+\partial_\rho \Gamma \kma \nn \\
{{\tilde\varphi}_{5}}^{~4} =\frac{(\partial_\rho+\Gamma+\pi_4)\xi}{2G}\kma~\quad
3{\varphi_4}^4+{\varphi_5}^5 = \xi - \Gamma \frac{(\partial_\rho+\Gamma+\pi_4 )\xi}{G} \kma  \\
{\varphi_4}^4-{\varphi_5}^5 = \frac{1}{2\Gamma}
(2\mu {{\tilde\varphi}_{5}}^{~4} - (\partial_\rho+\pi_4) (3{\varphi_4}^4+{\varphi_5}^5) ) \pkt \nn
\eea
After some awkward algebra, the Eq.~(\ref{scalar_eq_2}b) gives following for $\xi$:
\bea \label{seq}
k^2 \sigma^{-1} \xi = - \partial_\rho^2 \xi+ A(\rho) \partial_\rho \xi +B(\rho) \xi\kma \\
\label{ab}
A(\rho)=2\frac{\partial_\rho G}{G}-\Gamma\kma~\qquad
B(\rho)=\mu+\frac{1}{2}\pi_4^2+2\pi_4 \Gamma +2( \Gamma+\pi_4 ) \frac{\partial_\rho G}{G}\pkt
\eea
It is convenient to change to the dimensionless variable $x=\sqrt{\frac{5\Lambda}{8}} \rho$,
$x \in [0, \pi/2]$, in the Eqs.~(\ref{seq}, \ref{ab}).

Now we bring Schr\"odinger  Eq.(\ref{seq}) into standard form in two steps.
First we change independent variable from $x$ to $\tau$ according to $d\tau=dx/\sqrt{\sigma}$.
When $x \in [0, \pi/2]$ variable $\tau$ is changing from $0$ to $\tau_{max}$,
\be
\tau_{max}=\int^{\pi/2}_0 \frac{dx}{[cos(x)]^{2/5}}=\frac{\pi^{3/2} csc(\frac{3\pi}{10})}
{2 \Gamma (\frac{7}{10}) \Gamma (\frac{4}{5})} \approx 2.27221542 \pkt
\ee
This way we get rid of factor $\sigma$ in the l.h.s. of Eq.~(\ref{seq}) and coefficients $A$ and $B$
are changed to
\be
A_{\tau}=\sqrt{\sigma} A+\frac{1}{2}\sqrt{\sigma} \pi_4 \kma B_{\tau}=\sigma B \pkt
\ee
Next with the transformation $\xi = {\rm exp}(\frac{1}{2}\int A_{\tau}d\tau) \chi$
we get rid of the first derivative term and arrive at
\be \label{seq_standard}
(-\frac{d^2}{d\tau^2} + U) \chi = k^2 \chi \kma~\qquad {\rm with }~\qquad
U=B_{\tau}+\frac{1}{4} A_{\tau}^2 -\frac{1}{2} \partial_{\tau} A_{\tau} \pkt
\ee
Potential $U$ close to $\tau \to 0$ behaves as
\be \label{U1_at_zero}
U= \frac{35}{4 \tau^2} + O(\tau^2)\qquad {\rm for} \qquad m=1  \kma
\ee
and
\be \label{U_at_zero}
U= \frac{m^2-\frac{1}{4}}{\tau^2} +\frac{56-8 m^2}{15} + O(\tau^2) \qquad
{\rm for} \qquad m\geq 2   \kma
\ee
and close to $\tau_{max}$ as
\be \label{U_at_taumax}
U= -\frac{1}{4 (\tau_{max}-\tau)^2} \pkt
\ee
Note that quantum mechanical potential $U=-\gamma/x^2$ for $\gamma>\gamma_{cr}=1/4$
corresponds to unstable situation ("falling" to the center, see e.g. \cite{ll}).
So, in Eq.~(\ref{seq_standard}, \ref{U_at_taumax}) exactly critical case is realized,
which is on a border between stability and instability.
Another observation is that the $m=1$ case is distinguished, because potential $U$
is not negative in the inner region, while starting from $m=2$ it is negative not only
asymptotically $\tau \to \tau_{max}$, Eq.~(\ref{U_at_taumax}),
but also in the inner region, Fig.~1.

The regular branch of the wave function $\chi$ close to $\tau \to 0$ behaves as
\be
\chi \propto \tau^{7/2}-\frac{k^2}{16} \tau^{11/2}
\qquad {\rm for} \qquad  m=1  \kma
\ee
and
\be
\chi \propto \tau^{m+1/2}+\frac{56-8 m^2- 15 k^2}{60 (m+1)} \tau^{m+5/2}
\qquad {\rm for} \qquad  m\geq 2 \kma
\ee
and close to $\tau \to \tau_{max}$ as
\be
\chi \propto (\tau_{max}-\tau)^{1/2}  \pkt
\ee
To determine the number of bound states of Schr\"odinger equation
in a given potential we investigated the zero energy wave function.
According to known theorems (see e.g. \cite{aq95}) the number of nodes of zero energy wave
function exactly counts the number of negative energy states.
Solving numerically Schr\"odinger Eq.(\ref{seq_standard})
with above boundary conditions we found that there are no negative modes in $m=1$ case,
whereas there is single negative mode for each higher $m$.
We checked this statement up to $m=10$.

\section{Concluding remarks} \label{conclusions}
\par
We have shown that the solution described by Eqs.~(\ref{eq2},\ref{eq6},\ref{eq7},\ref{eq8}) and
corresponding to the $\Lambda_{\rm phys}=0$ is linearly unstable. The instability is related to
$\theta$-dependant perturbations, which are scalars under rotation of the four dimensional coordinates
$x_{\mu}$. We found that a single unstable mode appears in spectrum of linear perturbations
for each angular harmonic with $m\geq 2$.
A similar situation can be expected to arise for small $\Lambda_{\rm phys}$.
Even if the solution is stable starting from some $\Lambda^0_{\rm phys}\neq 0$,
it is quite improbable that the value of $\Lambda^0_{\rm phys}$ will be
$\sim 10^{-56} {\rm cm}^{-2}$, in agreement with current observations \cite{cc1}, \cite{cc2}.

Although the Rubakov-Shaposhnikov solution is found to be classically unstable, knowledge of unstable modes
can be useful, since it suggests the form of the stable solution to be sought.
Since the perturbations leading to instability are asymmetric under $\theta$-rotations,
it is clear that the initially symmetric state of the system tends to the more favored
asymmetric state. Lorentz invariance is not violated in the development of the instability,
as it would be if the instability  were related to the vector perturbations.
So, it is quite possible that the Eqs.~(\ref{eq4},\ref{eq5}) for $N=2$ have
asymmetric, $\theta$-dependent solution, which might be stable.

Since Rubakov-Shaposhnikov model is basic ingredient for many modern higher dimensional setups
it is natural to ask whether the instability disappears by adding extra fields.
So, question of stability should be carefully checked in each case.

\section*{Acknowledgements}
\par
I would like to thank Valery Rubakov for valuable comments on the manuscript.
The main part of work has been done during my visits to Geneva University, Switzerland and
to the Albert-Einstein-Institute, Golm, Germany.
I would like to thank the theory groups of these institutions and especially Ruth Durrer and
Hermann Nicolai for kind hospitality.
I also thank the Tomalla foundation and Georgian National Science Foundation
(Grant $\# GNSF/ST08/4-405$) for the financial support.



\begin{figure}[t]
\centerline{
\epsfig{file=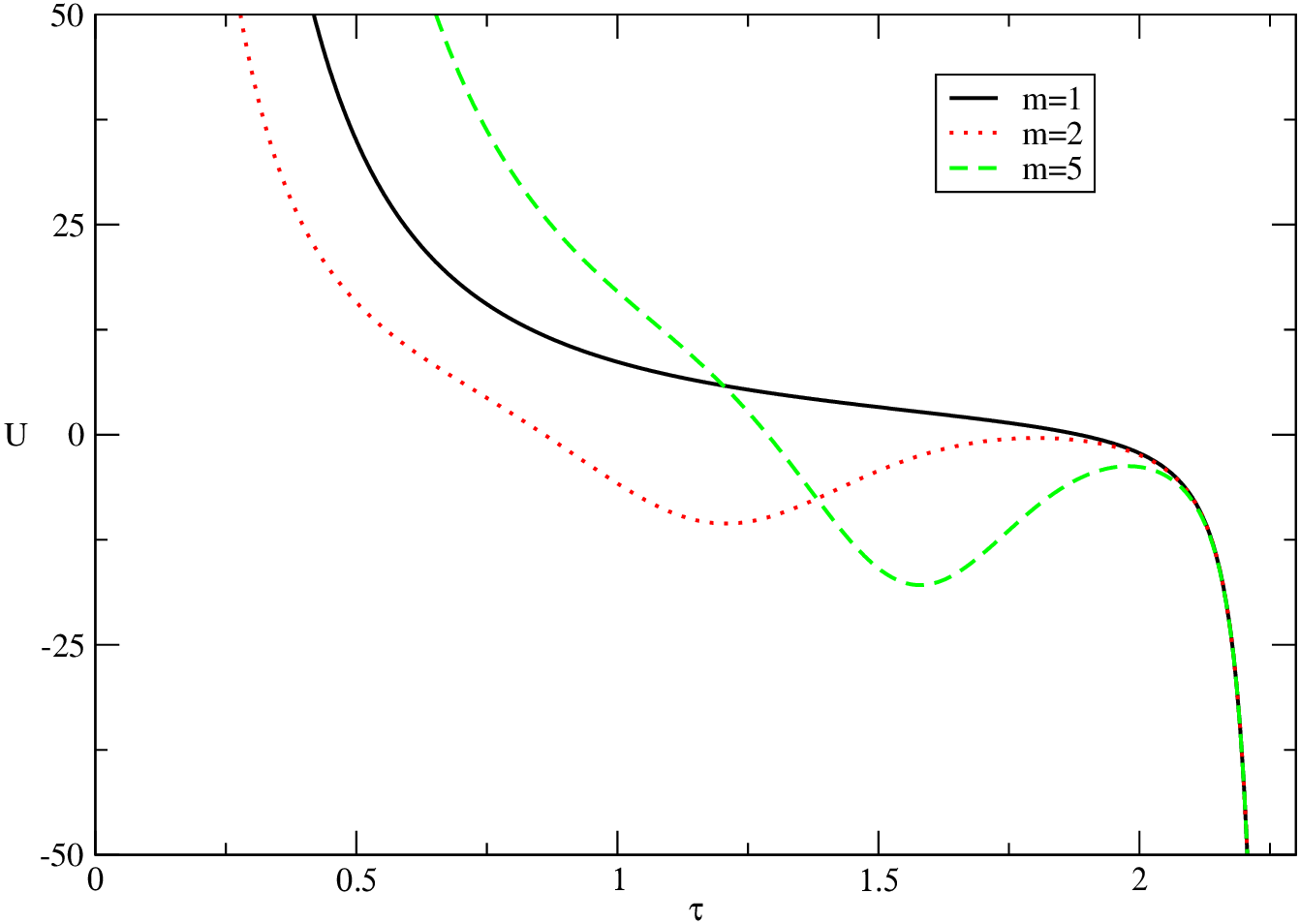,width=0.6\hsize,
bbllx=4.5cm,bblly=1.5cm,bburx=24.5cm,bbury=20.0cm}}
\labelcaption{pot}
{Shape of the potential $U(\tau)$ for $m=1,2$ and $5$.}
\end{figure}

\end{document}